\documentclass[12pt,a4paper,aps,prd,preprint,superscriptaddress,nofootinbib,preprintnumbers]{revtex4-1}
\usepackage[utf8]{inputenc}
\usepackage{graphicx,cancel,float}
\usepackage{subfigure}
\usepackage{amssymb}
\usepackage{textcomp}
\usepackage{amsmath}
\usepackage{tabularx}
\usepackage{bm}
\usepackage{times}
\usepackage{color}
\usepackage{slashed}
\usepackage{multirow}
\usepackage{verbatim}
\usepackage{epsfig,epstopdf}

\usepackage[colorlinks=true, pdfstartview=FitV, linkcolor=blue, citecolor=blue, urlcolor=blue]{hyperref}
\allowdisplaybreaks[4]

\newcommand{\minigraph}[5][0.25in]{\begin{minipage}{#2}\begin{center}\includegraphics[width=#2]{#5}\\\vspace{#3}\hspace{#1}{\footnotesize #4}\end{center}\end{minipage}}

\def\lsim{\mathrel{\raise.3ex\hbox{$<$\kern-.75em\lower1ex\hbox{$\sim$}}}}
\def\gsim{\mathrel{\raise.3ex\hbox{$>$\kern-.75em\lower1ex\hbox{$\sim$}}}}

\definecolor{orange}{rgb}{1,0.5,0}

\begin{document}

\title{Loop effect with vector mediator in the coherent neutrino-nucleus scattering}

\author{Wei Chao}
\email{chaowei@bnu.edu.cn}
\affiliation{
Center for Advanced Quantum Studies, Department of Physics,
Beijing Normal University, Beijing 100875, China
}
\author{Tong Li}
\email{litong@nankai.edu.cn}
\affiliation{
School of Physics, Nankai University, Tianjin 300071, China
}
\author{Jiajun Liao}
\email{liaojiajun@mail.sysu.edu.cn}
\affiliation{
School of Physics, Sun Yat-Sen University, Guangzhou 510275, China
}
\author{Min Su}
\email{201921140025@mail.bnu.edu.cn}
\affiliation{
Center for Advanced Quantum Studies, Department of Physics,
Beijing Normal University, Beijing 100875, China
}

\begin{abstract}
The observation of the coherent elastic neutrino-nucleus scattering (CE$\nu$NS) provides us opportunities to explore a wide class of new physics. In the Standard Model (SM), the CE$\nu$NS process arises from the vector and axial-vector neutral currents through the exchange of $Z$ boson and the axial-vector current contribution turns out to be subdominant. It is thus natural to consider the extra contributions to CE$\nu$NS from more generic new physics beyond the SM with (axial-)vector interactions associated with a new vector mediator $Z'$. Besides the ordinary CE$\nu$NS, the active neutrinos can convert into a new exotic fermion $\chi$ through the process $\nu N\to \chi N$ mediated by $Z'$ without violating the coherence. It would be interesting to consider the implication of this conversion for the new fermion sector beyond the SM. We consider the framework of a simplified neutrino model in which a new Dirac fermion $\chi$ interacts with active neutrinos and a leptophobic vector mediator $Z'$. We evaluate both the tree-level and loop-level contributions to the CE$\nu$NS and in particular the loop diagrams produce active neutrino elastic scattering process $\nu N\to \nu N$ with the fermion $\chi$ inside the loops.
When the interaction between $Z'$ and the SM quarks is vector type and axial-vector type, the CE$\nu$NS processes are respectively dominated by the tree-level and loop-level contributions. We investigate the constraints on the model parameters by fitting to the COHERENT data, assuming a wide range of $m_\chi$. The parameter space with $m_\chi$ larger than the maximal energy of incoming neutrinos can be constrained by including the loop-level contribution. More importantly, the inclusion of loop diagrams can place constraint on axial-vector interaction whose tree-level process is absent in the coherent neutrino-nucleus scattering.
\end{abstract}

\maketitle

\section{Introduction}
\label{sec:Intro}

The coherent elastic neutrino-nucleus scattering (CE$\nu$NS) process was first observed by
the COHERENT experiment~\cite{Akimov:2017ade} in the Spallation Neutron Source (SNS) at the Oak Ridge National Laboratory.
The neutrinos measured at COHERENT are produced by
the decays of stopped pion and muon, i.e. $\pi^+\to \mu^+~\nu_\mu$ and $\mu^+\to e^+~\nu_e~\bar{\nu}_\mu$. The energy of muon neutrinos is determined by $E_{\nu_\mu}=(m_\pi^2-m_\mu^2)/2m_\pi\simeq 30$ MeV and those of electron neutrinos and muon antineutrinos have the kinematic end point at $E_{\nu_e,\bar{\nu}_\mu}< m_\mu/2\simeq 53$ MeV.
For the neutrino-nucleus scattering, CE$\nu$NS occurs when the momentum transfer in the process is smaller than the inverse of the target nucleus radius.
The scattering amplitudes of the nucleons inside the nucleus can thus be summed all together coherently, which leads to a large enhancement of the cross section.
The CE$\nu$NS spectrum measured at COHERENT is consistent with the prediction of the Standard Model (SM), in which the CE$\nu$NS
process is generated through the weak neutral current~\cite{Freedman:1973yd}. Besides the
active neutrinos through $Z$ boson exchange in the SM, the CE$\nu$NS process could also produce an exotic fermion such as the right-handed (RH) neutrinos without violating the coherence condition. Thus, the COHERENT observation can provide us an opportunity to
explore the new physics (NP) associated with general neutrino interactions in the presence of
exotic fermion.

Recently, different groups studied the conversion from active neutrinos to an exotic fermion $\chi$ in the coherent neutrino-nucleus scattering~\cite{Brdar:2018qqj,Chang:2020jwl,Hurtado:2020vlj,Brdar:2020quo}
\begin{eqnarray}
\nu N\to \chi N\;.
\end{eqnarray}
For this inelastic scattering process with $E_\nu< m_\mu/2$, the kinematic constraint on the mass of the exotic fermion $\chi$ becomes $m_\chi<\sqrt{M(m_\mu+M)}-M\simeq 53$ MeV with $M$ being the nuclear mass~\cite{Brdar:2018qqj,Chang:2020jwl}. The COHERENT data can thus set bounds on this process only in the region of $m_\chi\lesssim 53$ MeV.
On the other hand, the validation of the coherence in CE$\nu$NS process depends on the specific interactions between neutrino and SM quark sector. Freedman et al. pointed out that the interactions such as axial quark current induce nuclear spin-dependent (SD) scattering and a cancellation between spin-up and spin-down nucleons~\cite{Freedman:1977xn}. They thus violate the coherence and the relevant CE$\nu$NS processes are suppressed for all nuclei except for light ones. The observation for heavy cesium-iodide (CsI) nuclei at COHERENT then can not place any constraint on the axial quark current and so on.

However, in the studies of the direct detection of WIMP dark matter (DM), it was emphasized that the interactions inducing SD scattering at tree-level can in turn generate spin-independent (SI) scattering through loop diagrams~\cite{Drees:1993bu,Hisano:2010ct,Baek:2016lnv,Baek:2017ykw,Arcadi:2017wqi,Li:2018qip,Abe:2018emu,Abe:2018bpo,Li:2019fnn,Mohan:2019zrk,
Ertas:2019dew,Giacchino:2015hvk,Giacchino:2014moa,Ibarra:2014qma,Colucci:2018vxz,Colucci:2018qml,Chao:2019lhb}. Moreover, the generated SI nuclear scattering cross section is independent of the momentum transfer $q\sim\mathcal{O}(\rm MeV)$ in the scattering and is not suppressed at leading order. The enhancement by the squared nuclear mass number in the coherent SI scattering compensates the suppression from the perturbative loop calculation. As a result, for the pseudoscalar or axial quark interaction, the full calculation involving the loop corrections would lead to sizable recoil events in DM direct detection experiments. We can apply the spirit of this loop effect to consider the loop corrections of the above inelastic neutrino scattering process~\cite{Li:2020pfy}. Besides the tree-level $\nu N\to \chi N$ process, the loop diagrams produce active neutrino elastic scattering process $\nu N\to \nu N$ with the above exotic fermion $\chi$ inside the loops. The involvement of loop diagrams would extend the constrained mass of fermion $\chi$ to $m_\chi>53$ MeV because there is no kinematic bound on the internal fermion $\chi$ in $\nu N\to \nu N$ process. More importantly, the loop diagrams induce non-momentum-suppressed CE$\nu$NS process and thus make the coupling of pseudoscalar or axial quark current become constrained by the COHERENT data. In this work we consider a simplified model of a fermionic particle $\chi$ interacting with SM neutrinos through a neutral vector boson $Z'$. The $Z'$ boson interacts with SM quarks in a general form and we discuss the pure vector or axial-vector current between $Z'$ and the SM quarks. The axial-vector interaction at tree-level in particular forbids the coherent neutrino-nucleus scattering. We investigate the SI elastic scattering generated by the loop corrections and the constraint on the model parameters by the COHERENT data.

This paper is organized as follows. In Sec.~\ref{sec:Model} we describe the simplified neutrino model with a vector mediator $Z'$. The Lagrangian is given in a general form with pure vector or axial-vector current for the interactions between $Z'$ and the SM quarks. In
Sec.~\ref{sec:CEnuNT} we present the analytical expressions of the CE$\nu$NS cross section. Both the tree-level and
loop-level contributions are given for the cases of vector and axial-vector currents. The numerical constraints on the couplings by the COHERENT data are also shown in Sec.~\ref{sec:Res}. Our conclusions are drawn in Sec.~\ref{sec:Con}.
The details of loop calculations are collected in the Appendix.

\section{Simplified neutrino model with a vector mediator}
\label{sec:Model}

The observation of the CE$\nu$NS process opens a wide investigation of physics opportunities. Due to the large uncertainties, possibly sizable contributions from new physics beyond the SM (BSM) can be constrained. In the SM, the CE$\nu$NS process arises from the neutral
current vector and axial-vector interactions through the exchange of $Z$ boson~\cite{Freedman:1973yd}, with the axial contribution being subdominant~\cite{Freedman:1977xn}. It is thus natural to consider to test more generic BSM physics with (axial-)vector interactions, such as neutrino-quark non-standard interactions (NSI) through (axial-)vector interaction as well as mediators associated with new $U(1)$ gauge symmetries. Suppose the SM particles are charged under a new $U(1)$ gauge symmetry, new vector bilinear couplings $Z'_\mu\bar{f}\gamma^\mu f$ ($f$ standing for the SM fermions) could arise from $U(1)_B$, $U(1)_{B-L}$, $U(1)_{B+L}$ or non-universal $U(1)_{B_i-L_j}$ models. If two chiral components of $f$ carry opposite $U(1)$ charges, the new gauge interaction would be axial-vector type, i.e. $Z'_\mu\bar{f}\gamma^\mu \gamma_5 f$. If only one certain chiral component of $f$ carries non-zero charge, new gauge interactions would be $Z'_\mu\bar{f}\gamma^\mu P_{L,R}f$.
On the other hand, it is interesting to explore the possibility of new fermion production through coherent scattering process.
This would make sense for exploring new fermion sector beyond the SM. The neutrinos can mix with a new fermion $\chi$ through a Yukawa interaction. As a result, the new fermion $\chi$ also interacts with neutrinos mediated by a vector $Z'$. Below we consider a model-independent way to generally study the conversion from active neutrinos to an exotic fermion $\chi$ via (axial-)vector interactions with SM quarks. Suppose the charged leptons are also charged under the new $U(1)$, the leptonic couplings also induce neutrino-nucleus scattering through kinetic mixing whose contribution is however highly suppressed. To study the relevant CE$\nu$NS process, we assume a leptophobic vector mediator for simplicity.

We study a simplified model of a Dirac fermionic particle $\chi$ interacting with SM neutrino and a neutral vector boson $Z'$ after the electroweak symmetry breaking. The Lagrangian is given by~\footnote{The ``inverse'' process can be induced by the same model, i.e. $\chi N\to \nu N$. If the fermion $\chi$ is long-lived enough, it serves as dark matter candidate and can be detected through this kind of fermionic absorption~\cite{Dror:2019onn,Dror:2019dib}, in which a single neutrino is produced in dark matter scattering. This process also leads to interesting
signals that can be searched for in dark matter experiments.}
\begin{eqnarray}
\mathcal{L}\supset Z'_\mu \bar{\chi}\gamma^\mu\Big(g_{\chi L}P_L + g_{\chi R}P_R \Big)\nu + Z'_\mu \sum_q \bar{q}\gamma^\mu\Big(g_{q L}P_L + g_{q R}P_R \Big) q  + h.c.\; ,
\end{eqnarray}
where the $Z'$ is assumed to be leptophobic and the quark couplings $g_{qL}, g_{qR}$ are not flavor universal. The couplings $g_{\chi L}, g_{\chi R}$ denote the generic mixing between $\chi$ and neutrinos.
The massive gauge field $Z'$ is associated with a pseudo-Nambu Goldstone boson (pNGB) $\varphi$.
The Yukawa interaction of fermions with $\varphi$ is given by~\cite{Lavoura:2003xp}
\begin{eqnarray}
\mathcal{L}\supset \varphi {-i\over m_{Z'}} m_\chi\bar{\chi}\Big(g_{\chi L}P_L+g_{\chi R}P_R \Big)\nu + \varphi {i\over m_{Z'}}m_q\bar{q}\Big((g_{qR}-g_{qL})P_L+(g_{qL}-g_{qR})P_R \Big)q +h.c.\;, \label{lagrang}
\end{eqnarray}
in the limit of massless neutrino. We use the Feynman-'tHooft gauge to calculate the diagrams in Fig.~\ref{fig:box}.
Apparently, the above simplified hypothesis does not respect gauge invariance before the SM electroweak symmetry breaking. A UV model can be realized through a broken new $U(1)$ symmetry above the electroweak scale and a mixing between the fermion $\chi$ and a neutrino~\cite{Dror:2019onn,Dror:2019dib}. In this scenario, only $\chi$ and the SM quarks are charged under the new $U(1)$ and meanwhile $\chi$ mixes with the neutrinos through a Yukawa interaction. In the following we utilize the above simplified model to exhibit the loop effect in the CE$\nu$NS without loss of generality.

We further define two scenarios of the quark couplings
\begin{itemize}
\item Case A: $g_{qL}=g_{qR}=g_q$
\item Case B: $g_{qL}=-g_{qR}=g_q$
\end{itemize}
and assume the couplings of all quark species are not universal. The choice of case A is exactly the case with pure vector interaction between $Z'$ and the SM quark sector. In this case the Yukawa interaction involving the pNGB $\varphi$ is absent. It leads to the SI neutrino-nucleus scattering $\nu N\to \chi N$ at tree-level and the COHERENT data can place constraint on the interaction for $m_\chi\lesssim 53$ MeV. By contrast, the case B induces the axial-vector interaction at tree-level, i.e. $Z'_{\mu}\bar{q}\gamma^\mu \gamma_5 q$, and the corresponding SD neutrino-nucleus scattering is absent in CE$\nu$NS process. The relevant tree-level inelastic scattering can not be constrained by COHERENT data. We next consider the loop corrections for the two scenarios and the $\nu N\to \nu N$ process constrained by COHERENT data.

\section{Coherent elastic neutrino-nucleus scattering}
\label{sec:CEnuNT}

\begin{figure}[h!]
\begin{center}
\minigraph{8cm}{-0.05in}{(a)}{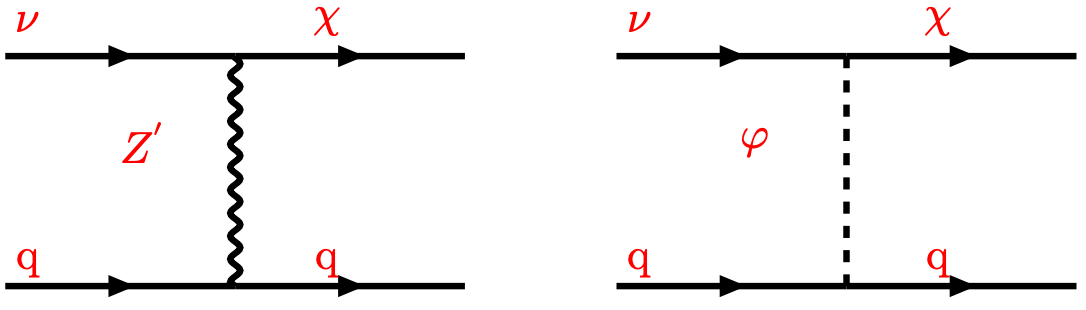}\\
\minigraph{8cm}{-0.05in}{(b)}{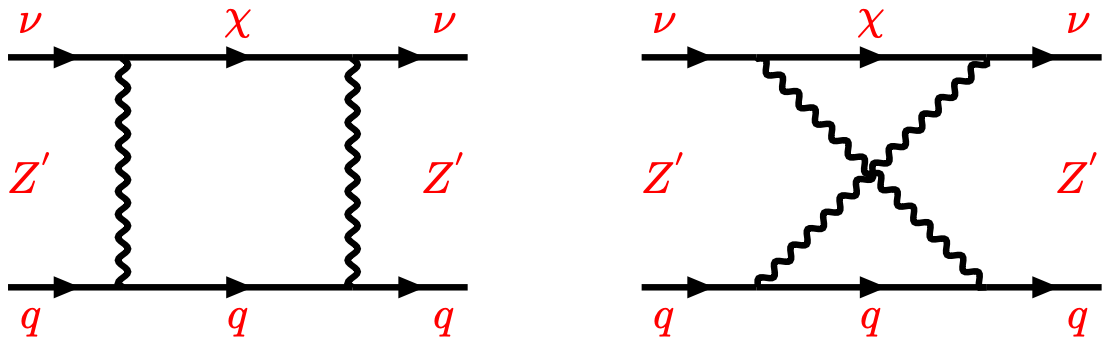}\\
\minigraph{8cm}{-0.05in}{(c)}{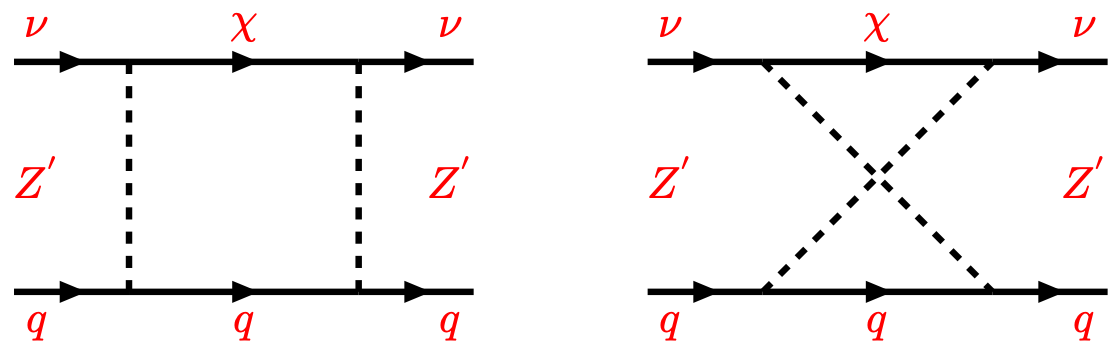}\\
\minigraph{8cm}{-0.05in}{(d)}{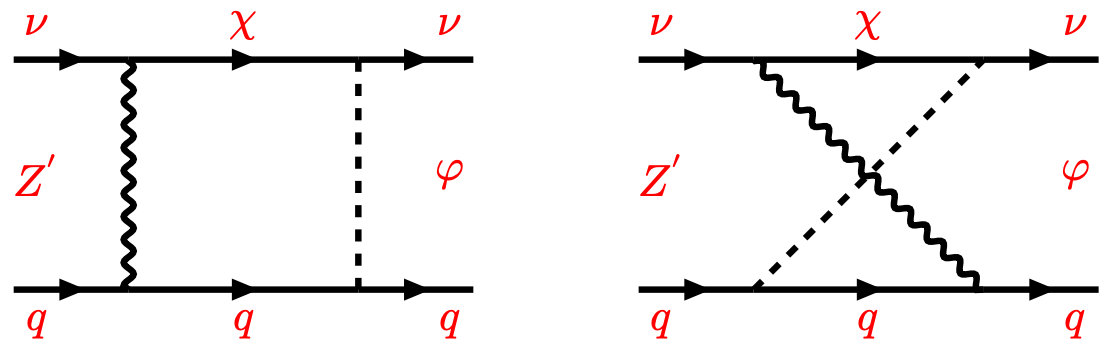}\\
\minigraph{8cm}{-0.05in}{(e)}{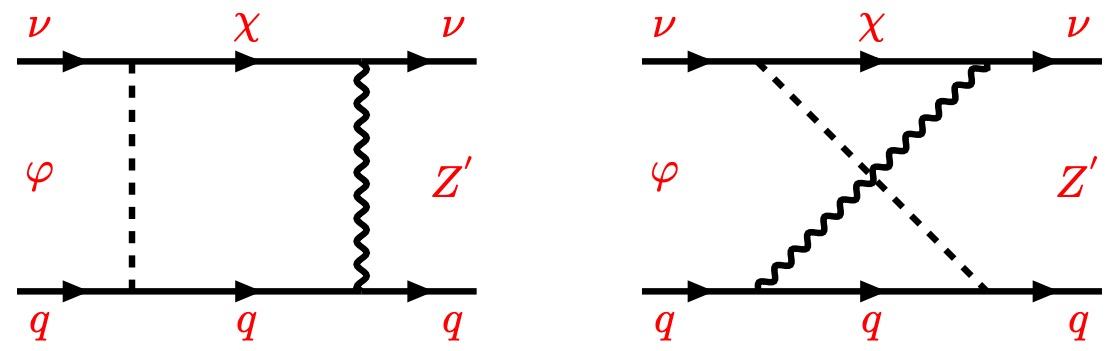}
\end{center}
\caption{Diagrams for the scattering processes of neutrino, (a): tree-level diagrams, (b): box diagrams with $Z'$ as mediator, (c): box diagrams with the pNGB $\varphi$ as mediator, (d) and (e): box diagrams with $Z'$-$\varphi$ as mediator.}
\label{fig:box}
\end{figure}

In this section we evaluate the coherent neutrino-nucleus scattering induced by the two coupling choices at both tree-level and loop-level. We first derive the scattering matrix elements at quark level.
For $\nu(p_1) q(k_1)\to \chi(p_2) q(k_2)$ as shown in Fig.~\ref{fig:box} (a), we obtain the general tree-level matrix element
\begin{eqnarray}
i\mathcal{M}_{\rm tree}&=&\sum_{q=all} {i\over t-m_{Z'}^2} \bar{\chi}(p_2) \gamma_\mu g_{\chi L} P_L \nu(p_1) \bar{q}(k_2)\gamma^\mu (g_{qL}P_L + g_{qR}P_R) q(k_1)\nonumber \\
&+&\sum_{q=all}{-i m_\chi m_q\over m_{Z'}^2} \bar{\chi}(p_2) g_{\chi L} P_L \nu(p_1) \bar{q}(k_2) \Big[ (g_{qR}-g_{qL})P_L + (g_{qL}-g_{qR})P_R \Big] q(k_1) \;,
\end{eqnarray}
with the Mandelstam variable $t=(p_1-p_2)^2$. For case A the first term induces vector current $\bar{q}\gamma^\mu q$ and the second term vanishes. The two terms result in either axial-vector or pseudo-scalar current in case B.

The elastic scattering process $\nu(p_1)q(k_1)\to\nu(p_2)q(k_2)$ occurs through multiple loop-level contributions to the leading SI effective operators. The SI effective operators
decompose into the scalar operator $\bar{\nu}\nu\bar{q}q$, the twist-2 neutrino-quark operators and the neutrino-gluon scalar operator $\bar{\nu}\nu G^a_{\mu\nu} G^{a\mu\nu}$. All of the one-loop box diagrams in Fig.~\ref{fig:box} contribute to the scalar operator with each of the light quarks $u,d,s$ as well as the twist-2 operators for $q=u,d,s,c,b$. The cases A and B both have one-loop box diagrams with the vector mediator $Z'$ as shown in Fig.~\ref{fig:box} (b)
\begin{eqnarray}
&&i{\cal M}_{{\rm box}-Z'Z'}=\nonumber \\
&&-\sum_{q=u,d,s}\frac{i}{(4\pi)^2}g_{\chi L}g_{\chi R} g_{q}^2m_\chi m_q \Big[\kappa D_{0}^b+\frac{8}{m_{Z'}^2}(D_{00}^a-D_{00}^b)\Big]\bar{\nu}(p_2) P_L \nu(p_1)\bar{q}(k_2)q(k_1)\nonumber \\
&&+\sum_{q=u,d,s,c,b}\frac{i}{(4\pi)^2}g_{\chi L}^2 g_{q}^2\Big[\frac{16}{m_{Z'}^2}(4D_{001}^a-4D_{001}^b+D_{00}^a-D_{00}^b)+ 8 (D_1^b+D_0^b)\Big]\bar{\nu}(p_2)i\partial^\mu \gamma^\nu P_L\nu(p_1){\cal O}_{\mu\nu}^q\nonumber\\
&&-\sum_{q=u,d,s,c,b}\frac{i}{(4\pi)^2}g_{\chi L}g_{\chi R} g_{q}^2\frac{8}{m_{Z'}^2}m_\chi (D_{11}^a-D_{11}^b)\bar{\nu}(p_2)i\partial^\mu i\partial^\nu P_L\nu(p_1){\cal O}_{\mu\nu}^q\;,
\end{eqnarray}
where $\kappa=-4~(12)$ for case A (B) is a pre-factor in front of the loop function $D_0^b$ obtained after the Lorentz contraction, $p_1, p_2 (k_1, k_2)$ denote the momenta of neutrinos (quarks), and ${\cal O}_{\mu\nu}^q$ is the twist-2 operator for quark
\begin{eqnarray}
{\cal O}_{\mu\nu}^q={i\over 2}\bar{q}\Big(\partial_\mu\gamma_\nu+\partial_\nu\gamma_\mu-{1\over 2}g_{\mu\nu}\cancel{\partial}\Big)q\;.
\end{eqnarray}
The Passarino-Veltman functions used here and below are collected in the Appendix.
The case B has additional contributions involving the pNGB $\varphi$ as shown in Figs.~\ref{fig:box} (c), (d) and (e). The matrix element of the one-loop box diagrams with the would-be Goldstone mediator $\varphi$ is
\begin{eqnarray}
i{\cal M}_{{\rm box}-\varphi\varphi}({\rm B})&=&
-\sum_{q=u,d,s}{i\over(4\pi)^2}g_{q}^2 g_{\chi L}g_{\chi R}{32 m_\chi^3 m_q^3 \over m_{Z'}^6}(D_{00}^a-D_{00}^b)\bar\nu(p_2)P_L\nu(p_1) \bar q(k_2)q(k_1)\nonumber\\
&&-\sum_{q=u,d,s,c,b}{i\over(4\pi)^2}g_{q}^2 g_{\chi L}^2{64 m_\chi^2 m_q^2 \over m_{Z'}^6}(D_{001}^a-D_{001}^b)\bar\nu(p_2) i\partial^\mu \gamma^\nu P_L\nu(p_1){\cal O}_{\mu\nu}^q\nonumber\\
&&-\sum_{q=u,d,s,c,b}{i\over(4\pi)^2}g_{q}^2 g_{\chi L}g_{\chi R}{32 m_\chi^3 m_q^2 \over m_{Z'}^6}(D_{11}^a-D_{11}^b)\bar\nu(p_2) i\partial^\mu i\partial^\nu P_L\nu(p_1){\cal O}_{\mu\nu}^q\;.
\end{eqnarray}
The one-loop box diagrams of the $Z'$-$\varphi$ mixing mediator induce the following matrix element
\begin{eqnarray}
i{\cal M}_{{\rm box}-Z'\varphi}({\rm B})=\sum_{q=u,d,s}{i\over(4\pi)^2}g_q^2 g_{\chi L}g_{\chi R}{32 m_\chi m_q \over m_{Z'}^2}D_{00}^b \bar\nu(p_2) P_L\nu(p_1)\bar q(k_2) q(k_1)\;.
\end{eqnarray}

\begin{figure}[htb!]
\begin{center}
\includegraphics[scale=1,width=0.7\linewidth]{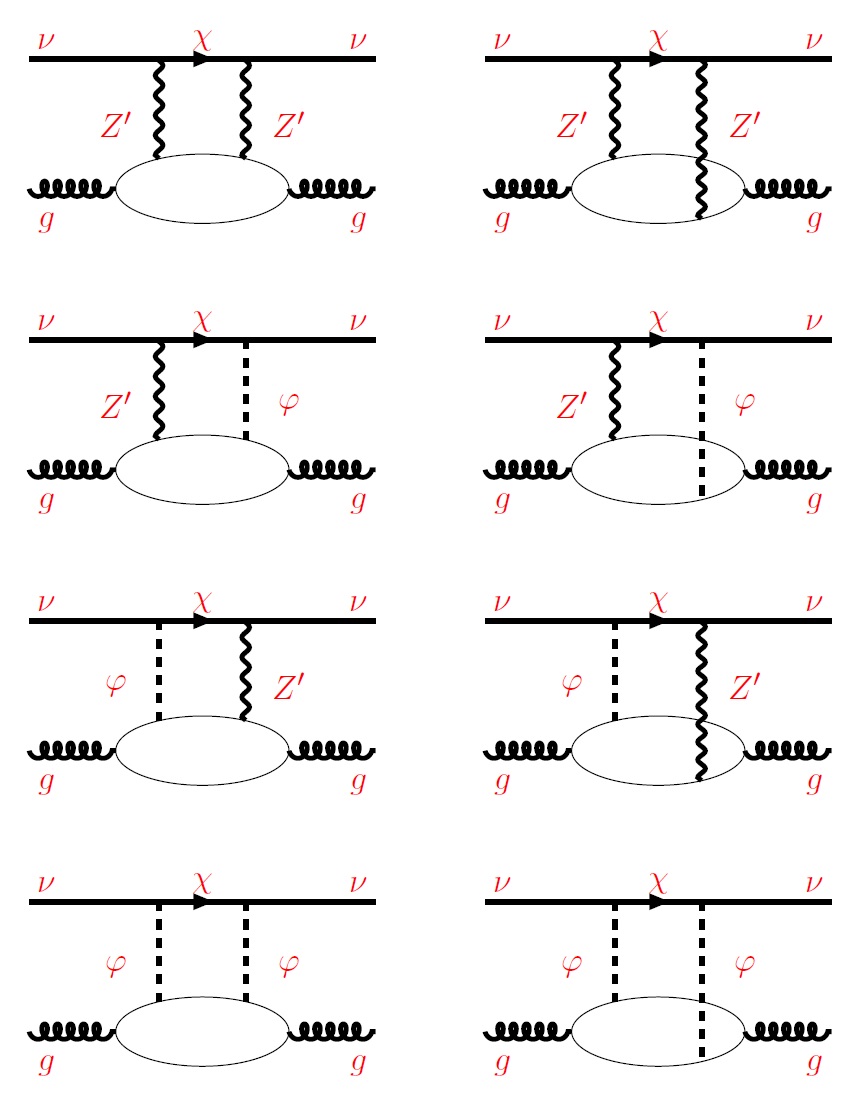}
\end{center}
\caption{Two-loop diagrams for effective gluon interactions with $Z'$ and pNGB $\varphi$ mediators.}
\label{fig:2loop}
\end{figure}

The contributions of heavy quarks should be integrated out at the two-loop level and cause the effective interaction between neutrino and gluon fields with the same order of magnitude.
For heavy quarks running in two-loop diagrams in Fig.~\ref{fig:2loop}, we calculate the amplitude using the Fock-Schwinger gauge~\cite{Shifman:1980ui,Novikov:1983gd,Kummer:1985xr,Delbourgo:1991pn} for the gluon background field in zero momentum limit~\cite{Hisano:2010ct,Abe:2015rja,Hisano:2017jmz,Abe:2018emu}.
The complete two-loop matrix elements with two $Z^\prime$ running in the loop in the Fig.~\ref{fig:2loop} read as follows
\begin{eqnarray}
i{\cal M}_{{\rm 2loop}-Z'Z'}&=&
\kappa'\sum_{q=c,b}{i\over(4\pi)^2}{\alpha_s\over 12\pi}G_{\alpha\beta}^a G^{a\alpha\beta} g_{q}^2g_{\chi L}g_{\chi R} m_\chi D_0^b\bar\nu(p_2)P_L\nu(p_1)\;,
\end{eqnarray}
where $G^a_{\alpha\beta}$ is the gluon field strength tensor and $\kappa'=-3~(13)$ for case A (B). The contribution from the top quark loop is suppressed by the top quark mass and we neglect its contribution in our numerical calculation.
For case B, there is additional contribution from two pNGB mediators in the loop
\begin{equation}
i{\cal M}_{{\rm 2loop}-\varphi\varphi}({\rm B})=-\sum_{q=c,b,t}{i\over(4\pi)^2}{\alpha_s\over 12\pi}G_{\alpha\beta}^a G^{a\alpha\beta} {2g_q^2 m_q^2 m_\chi^3 \over m_{Z'}^4}g_{\chi L}g_{\chi R} {\partial \over \partial m_{Z'}^2} F_G(m_{Z'}^2)\bar\nu(p_2)P_L\nu(p_1)\;,
\end{equation}
where the $F_G$ function is also collected in Appendix. As we adopt downphilic coupling $g_q$, in the numerical calculation, we neglect the two-loop diagrams dominated by top quark contribution.
For $Z'$-$\varphi$ mixing mediators, the two-loop contribution vanishes.

Next, we can obtain the matrix elements at nucleon-level in terms of the nucleon form factors. The nucleon form factors are defined as~\cite{DelNobile:2013sia,Bishara:2017pfq}
\begin{eqnarray}
\langle N| m_q \bar{q}q|N\rangle&=&m_N f_q^N \bar{N}N\;, \quad q=u,d,s\;,\\
\langle N| m_Q \bar{Q}Q|N\rangle&=&\langle N| {-\alpha_s\over 12\pi} G^a_{\mu\nu}G^{a\mu\nu}|N\rangle = {2\over 27} m_N f_G^N \bar{N}N\;, \quad Q=c,b,t\;,\\
\langle N| {\cal O}_{\mu\nu}^q|N\rangle&=& {1\over m_N} \Big(p^N_\mu p^N_\nu -{1\over 4}m_N^2 g_{\mu\nu} \Big)\Big(q^N(2)+\bar{q}^N(2)\Big) \bar{N}N\;, \quad q=u,d,s,c,b\;,\\
\langle N| \bar{q}\gamma^\mu q|N\rangle&=& c_q^N \bar{N}\gamma^\mu N\;, \quad c_u^p = c_d^n = 2, c_d^p = c_u^n = 1, c_s^p=c_s^n=0\;,
\end{eqnarray}
for the SI interactions. The nucleon-level tree diagram of $\nu N\to \chi N$ then becomes
\begin{eqnarray}\label{eq:treeM}
i\mathcal{M}_{\rm tree}^N
&=&\sum_{q=u,d,s} {i\over t-m_{Z'}^2}{g_{\chi L}(g_{qL}+g_{qR})\over 2}c_q^N \bar{\chi}(p_2) \gamma_\mu P_L \nu(p_1)\bar{N}(k_2)\gamma^\mu N(k_1)+\fbox{SD}\nonumber \\
&=&\left\{
     \begin{array}{ll}
       \sum_{q=u,d,s}c_q^N {i\over t-m_{Z'}^2}g_{\chi L}g_q \bar{\chi}(p_2) \gamma_\mu P_L \nu(p_1)\bar{N}(k_2)\gamma^\mu N(k_1), & \hbox{case A} \\
       0, & \hbox{case B}
     \end{array}
   \right.
+\fbox{SD}\;,
\end{eqnarray}
where $\fbox{SD}$ stands for SD terms which will be omitted in the following calculation~\footnote{In principle, the SD terms also contribute to the CE$\nu$NS process~\cite{Hoferichter:2020osn}. As it is highly suppressed compared with the SI terms, we will not consider their contribution here.}.
The nucleon-level box diagrams for $\nu N\to \nu N$ are
\begin{eqnarray}
i{\cal M}_{{\rm box}-Z'Z'}^N &=& -\frac{i}{(4\pi)^2}g_{\chi L}g_{\chi R} g_{q}^2 m_\chi\bar{\nu}(p_2) P_L \nu(p_1)\bar{N}(k_2)N(k_1) \nonumber \\
&&\times\Big[\kappa D_{0}^b+\frac{8}{m_{Z'}^2}(D_{00}^a-D_{00}^b)\Big]\Big(\sum_{q=u,d,s}m_N f_q^N\Big) \nonumber\\
&&+\frac{i}{(4\pi)^2}g_{\chi L}^2 g_{q}^2{(k_1\cdot p_1)\over m_N}\bar{\nu}(p_2)\slashed{k_1} P_L \nu(p_1)\bar{N}(k_2)N(k_1)\sum_{q=u,d,s,c,b}(q^N(2)+\bar{q}^N(2)) \nonumber\\
&&\times\Big[\frac{16}{m_{Z'}^2}(4D_{001}^a-4D_{001}^b+D_{00}^a-D_{00}^b)+ 8( D_1^b+D_{0}^b)\Big] \nonumber\\
&&-\frac{i}{(4\pi)^2}g_{\chi L}g_{\chi R} g_{q}^2\frac{8}{m_{Z'}^2}m_\chi{(k_1\cdot p_1)^2\over m_N}\bar{\nu}(p_2) P_L \nu(p_1)\bar{N}(k_2)N(k_1) \nonumber\\
&&\times(D_{11}^a-D_{11}^b)\sum_{q=u,d,s,c,b}(q^N(2)+\bar{q}^N(2))\;,
\label{eq:MZZ}
\end{eqnarray}
\begin{eqnarray}
\label{eq:MZphi}
i{\cal M}_{{\rm box}-Z'\varphi}^N({\rm B})&=&{i\over(4\pi)^2}g_q^2 g_{\chi L}g_{\chi R}{32m_\chi\over m_{Z'}^2}D_{00}^b\Big(\sum_{q=u,d,s}m_N f_q^N\Big)\bar\nu(p_2) P_L\nu(p_1)\bar N(k_2) N(k_1)\;,
\end{eqnarray}
\begin{eqnarray}
\label{eq:Mphiphi}
i{\cal M}_{{\rm box}-\varphi\varphi}^N({\rm B})&=&-{i\over(4\pi)^2}g_{q}^2 g_{\chi L}g_{\chi R}{32 m_\chi^3 \over m_{Z'}^6}\bar\nu(p_2)P_L\nu(p_1) \bar N(k_2)N(k_1)\nonumber\\
&&\times(D_{00}^a-D_{00}^b)\Big(\sum_{q=u,d,s}m_N m_q^2 f_q^N\Big)\nonumber\\
&&-{i\over(4\pi)^2}g_{q}^2 g_{\chi L}^2{64 m_\chi^2 \over m_{Z'}^6}{(k_1\cdot p_1)\over m_N}\bar{\nu}(p_2)\slashed{k_1} P_L \nu(p_1)\bar{N}(k_2)N(k_1)\nonumber\\
&&\times(D_{001}^a-D_{001}^b)\sum_{q=u,d,s,c,b}m_q^2 (q^N(2)+\bar{q}^N(2))\nonumber\\
&&-{i\over(4\pi)^2}g_{q}^2 g_{\chi L}g_{\chi R}{32 m_\chi^3 \over m_{Z'}^6}{(k_1\cdot p_1)^2\over m_N}\bar{\nu}(p_2) P_L \nu(p_1)\bar{N}(k_2)N(k_1)\nonumber\\
&&\times(D_{11}^a-D_{11}^b)\sum_{q=u,d,s,c,b}m_q^2 (q^N(2)+\bar{q}^N(2))\;.
\end{eqnarray}
The non-vanishing two-loop matrix elements are given by
\begin{eqnarray}
\label{eq:M2ZZ}
i{\cal M}_{{\rm 2loop}-Z'Z'}^N&=&-\kappa'\sum_{q=c,b}{i\over(4\pi)^2} g_{q}^2 g_{\chi L}g_{\chi R}m_\chi{2\over 27 }m_N f_G^N D_0^b \bar\nu(p_2)P_L\nu(p_1)\bar N(k_2)N(k_1)\;,
\end{eqnarray}
with $\kappa'=-3~(13)$ for case A (B) and
\begin{eqnarray}
\label{eq:M2phiphi}
i{\cal M}_{{\rm 2loop}-\varphi\varphi}^N({\rm B})&=&\sum_{q=c,b,t}{i\over(4\pi)^2}{2g_q^2 m_q^2m_\chi^3 \over m_{Z'}^4}g_{\chi L}g_{\chi R} {2\over 27 }m_N f_G^N F_G(p_1^2,m_\chi^2,m_{Z'}^2,m_q^2)\\\nonumber
&&\times \bar\nu(p_2) P_L\nu(p_1)\bar N(k_2) N(k_1)\;.
\end{eqnarray}

Since only the nuclear recoil is detected in a neutrino scattering experiment, the total differential cross section of CE$\nu$NS can be written as
\begin{align}
\frac{d\sigma}{dT}=\frac{d\sigma_\text{SM}}{dT}+\frac{d\sigma_\text{tree}}{dT}+\frac{d\sigma_\text{loop}}{dT}\,,
\end{align}
where $T$ denotes the nuclear recoil energy.
The differential cross section in the SM is given by
\begin{align}
\frac{d\sigma_\text{SM}}{dT}=\frac{G_F^2 M}{2\pi}[Z g_p^V + N g_n^V]^2F^2(Q^2)\left(2-\frac{MT}{E^2}-\frac{2T}{E}+\frac{T^2}{E^2}\right)\,,
\end{align}
where $E$ denotes the incoming neutrino energy, $Z$ ($N$) is the number of protons (neutrons) in the target nucleus, $g_n^V=-\frac{1}{2}$ and $g_p^V=\frac{1}{2}-2\sin^2\theta_W$ are the SM weak couplings with $\theta_W$ being the weak mixing angle. We neglect the radiative corrections of the SM weak couplings and only adopt the above values at tree-level.
Here $F(Q^2)$ represents the nuclear form factor with the moment transfer $Q^2=2MT$.
Since different form factor parameterizations have negligible effect on the COHERENT spectrum~\cite{Cadeddu:2017etk, AristizabalSierra:2019zmy}, we take the Helm parameterization~\cite{Helm:1956zz} for the nuclear form factor in our analysis.
Also, we use a single nuclear form factor for both the proton and neutron for simplicity. A more precise treatment would include different form factors for the proton and neutron considering the large uncertainty for the neutron form factor, which is driven by the poorly known root-mean-square radius of the neutron density distribution~\cite{ AristizabalSierra:2019zmy}.

From Eq.~(\ref{eq:treeM}), the tree-level differential cross section of $\nu N\to \chi N$ for case A is
\begin{align}
\frac{d\sigma_\text{tree}}{dT}&=\frac{g_{\chi L}^2 M F^2(Q^2)}{4\pi  (m_{Z'}^2+2MT)^2 }\Big[(2g_u+g_d)Z+(g_u+2g_d)N\Big]^2\nonumber \\
&\left[\left(2-\frac{MT}{E^2}-\frac{2T}{E}+\frac{T^2}{E^2}\right)-\frac{m_\chi^2}{2E^2}\left(1+2\frac{E}{M}-\frac{T}{M}\right)\right]\,.
\label{eq:treelevel}
\end{align}
For case B, the tree-level differential cross section of $\nu N\to \chi N$ is zero.
Note that in order to produce a massive fermion $\chi$ in the scattering $\nu N\to \chi N$, the energy of the incident neutrinos should be larger than a minimal energy~\cite{Brdar:2018qqj, Chang:2020jwl}, i.e.
\begin{align}
E>m_\chi+\frac{m_\chi^2}{2M}\,.
\end{align}
From Eqs.~(\ref{eq:MZZ}),~(\ref{eq:MZphi}),~(\ref{eq:Mphiphi}),~(\ref{eq:M2ZZ}) and~(\ref{eq:M2phiphi}),
we can write the loop-level differential cross section of $\nu N\to \nu N$ as
\begin{align}
\frac{d\sigma_\text{loop}}{dT}=\frac{M^3}{8\pi }F^2(Q^2)\left( 2M+ T \right)\left[T\left(\frac{c_1}{M E}+c_3 M E\right)^2+c_2^2\left(2ME(E-T)-M^2T \right)\right]\,,
\end{align}
where
\begin{align}
c_1=&-\frac{1}{(4\pi)^2}g_{\chi L}g_{\chi R}  m_\chi\Big[\kappa D_{0}^b+\frac{8}{m_{Z'}^2}(D_{00}^a-D_{00}^b)\Big]\sum_{q=u,d,s}g_{q}^2\Big(Zm_p  f_q^p+N m_n  f_q^n\Big) \nonumber\\
&-\kappa'{1\over(4\pi)^2}  g_{\chi L}g_{\chi R}m_\chi{2\over 27 } D_0^b\sum_{q=c,b}g_{q}^2\Big(Zm_p f_G^p+Nm_n f_G^n\Big)\,,\\
c_2=&\frac{1}{(4\pi)^2 }g_{\chi L}^2 \Big[\frac{16}{m_{Z'}^2}(4D_{001}^a-4D_{001}^b+D_{00}^a-D_{00}^b)+ 8( D_1^b+D_{0}^b)\Big]
\nonumber\\
\,\,&\times\sum_{q=u,d,s,c,b}g_{q}^2\Big[\frac{Z}{m_p}(q^p(2)+\bar{q}^p(2))+\frac{N}{m_n}(q^n(2)+\bar{q}^n(2))\Big]\,,\\
c_3=&-\frac{1}{(4\pi)^2}g_{\chi L}g_{\chi R} \frac{8m_\chi}{m_{Z'}^2}
(D_{11}^a-D_{11}^b)\sum_{q=u,d,s,c,b}g_{q}^2\Big[\frac{Z}{m_p}(q^p(2)+\bar{q}^p(2))+\frac{N}{m_n}(q^n(2)+\bar{q}^n(2))\Big]\,,
\end{align}
for case A, and
\begin{align}
c_1=&-\frac{1}{(4\pi)^2}g_{\chi L}g_{\chi R}  m_\chi\Big[\kappa D_{0}^b+\frac{8}{m_{Z'}^2}(D_{00}^a-D_{00}^b)\Big]\sum_{q=u,d,s}g_{q}^2\Big(Zm_p f_q^p+Nm_n f_q^n\Big)
\nonumber\\
&+ {1\over(4\pi)^2} g_{\chi L}g_{\chi R}{32m_\chi\over m_{Z'}^2}D_{00}^b\sum_{q=u,d,s}g_{q}^2\Big(Zm_p f_q^p+Nm_n f_q^n\Big)
\nonumber\\
&-{1\over(4\pi)^2} g_{\chi L}g_{\chi R}{32 m_\chi^3 \over m_{Z'}^6}(D_{00}^a-D_{00}^b)\sum_{q=u,d,s}g_{q}^2m_q^2\Big(Zm_p  f_q^p+Nm_n  f_q^n\Big)
\nonumber\\
&-\kappa'{1\over(4\pi)^2}  g_{\chi L}g_{\chi R}m_\chi{2\over 27 } D_0^b \sum_{q=c,b}g_{q}^2\Big(Zm_p f_G^p+Nm_n f_G^n\Big)
\nonumber\\
&+{1\over(4\pi)^2}g_{\chi L}g_{\chi R}{m_\chi^3 \over m_{Z'}^4} {4\over 27 } \sum_{q=c,b,t}g_q^2 m_q^2F_G(p_1^2,m_\chi^2,m_{Z'}^2,m_q^2)\Big(Zm_p f_G^p+Nm_n f_G^n\Big)\,,\\
c_2=&\frac{1}{(4\pi)^2 }g_{\chi L}^2 \Big[\frac{16}{m_{Z'}^2}(4D_{001}^a-4D_{001}^b+D_{00}^a-D_{00}^b)+ 8( D_1^b+D_{0}^b)\Big]
\nonumber\\
&\times\sum_{q=u,d,s,c,b}g_{q}^2\Big[\frac{Z}{m_p}(q^p(2)+\bar{q}^p(2))+\frac{N}{m_n}(q^n(2)+\bar{q}^n(2))\Big]
\nonumber\\
&-{1\over(4\pi)^2} g_{\chi L}^2{64 m_\chi^2 \over m_{Z'}^6 }(D_{001}^a-D_{001}^b)
\nonumber\\
&\times\sum_{q=u,d,s,c,b}g_{q}^2m_q^2\Big[\frac{Z}{m_p}(q^p(2)+\bar{q}^p(2))+\frac{N}{m_n}(q^n(2)+\bar{q}^n(2))\Big]\,,\\
c_3=&-\frac{1}{(4\pi)^2}g_{\chi L}g_{\chi R} \frac{8m_\chi}{m_{Z'}^2}(D_{11}^a-D_{11}^b)
\nonumber\\
&\times\sum_{q=u,d,s,c,b}\Big[\frac{Z}{m_p}(q^p(2)+\bar{q}^p(2))+\frac{N}{m_n}(q^n(2)+\bar{q}^n(2))\Big]
\nonumber\\
&-{1\over(4\pi)^2} g_{\chi L}g_{\chi R}{32 m_\chi^3 \over m_{Z'}^6 }(D_{11}^a-D_{11}^b)
\nonumber\\
&\times\sum_{q=u,d,s,c,b}g_{q}^2m_q^2\Big[\frac{Z}{m_p}(q^p(2)+\bar{q}^p(2))+\frac{N}{m_n}(q^n(2)+\bar{q}^n(2))\Big]\,,
\end{align}
for case B. One can see that the $c_1$ coefficients are from both quark and gluon scalar operators. The coefficients $c_2,c_3$ are only dependent on the twist-2 operator. For the nucleon form factors in SI interactions, we adopt the default values in micrOMEGAs~\cite{Belanger:2008sj,Belanger:2018ccd}.

\begin{figure}[htb!]
\begin{center}
\includegraphics[scale=1,width=0.65\linewidth]{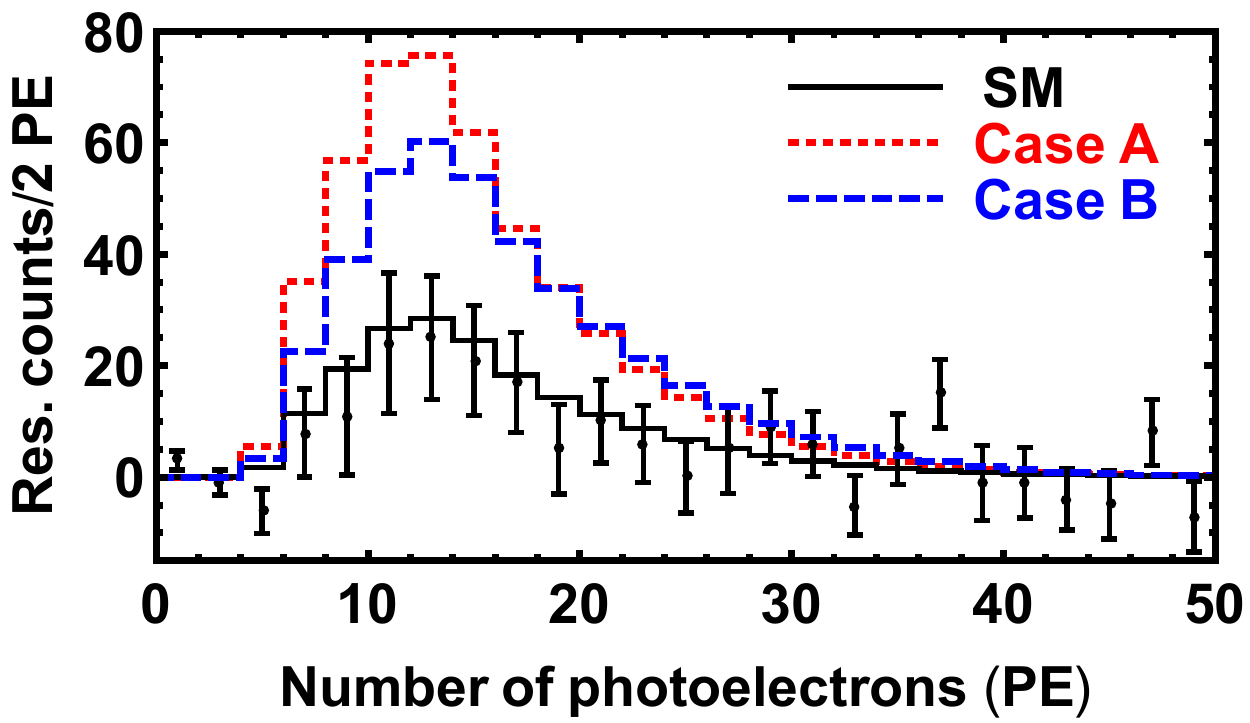}
\end{center}
\caption{
The expected CE$\nu$NS residual event as a function of the number of photoelectrons at COHERENT. The black solid lines correspond to the SM case, and the red dotted (blue dashed) lines correspond to Case A (B) with  $m_\chi=10$ MeV, $m_{Z^\prime}=100$ MeV, and $g_\chi g_q=5.0\times10^{-8}$ ( $m_\chi=100$ MeV, $m_{Z^\prime}=1000$ MeV, and $g_\chi g_q=5.0\times10^{-4}$). Here we assume $g_{\chi L}=g_{\chi R}=g_{\chi}$.
}
\label{fig:spectrum}
\end{figure}

The neutrinos measured at COHERENT are generated from the stopped pion decays and the muon decays, and their fluxes are given by
\begin{align}
\label{eq:nu-spectra.COHERENT}
\phi_{\nu_\mu}(E_{\nu_\mu})&= \mathcal{N}_0
\frac{2m_\pi}{m_\pi^2-m_\mu^2}\,
\delta\left(
1-\frac{2E_{\nu_\mu}m_\pi}{m_\pi^2-m_\mu^2}
\right) \ ,
\nonumber\\
\phi_{\nu_e}(E_{\nu_e})&= \mathcal{N}_0 \frac{192}{m_\mu}
\left(\frac{E_{\nu_e}}{m_\mu}\right)^2
\left(\frac{1}{2}-\frac{E_{\nu_e}}{m_\mu}\right)\ ,\\
\phi_{\overline\nu_\mu}(E_{\overline\nu_\mu})&= \mathcal{N}_0  \frac{64}{m_\mu}
\left(\frac{E_{\overline\nu_\mu}}{m_\mu}\right)^2
\left(\frac{3}{4}-\frac{E_{\overline\nu_\mu}}{m_\mu}\right)\,,\nonumber
\end{align}
where the normalization factor $\mathcal{N}_0 =\frac{rtN_\text{POT}}{4\pi L^2}$ with $r=0.08$ being the number of neutrinos per flavor produced per proton collision, $t$ the number of years of data collection, $N_\text{POT}=2.1\times 10^{23}$ the total number of protons on target per year, and $L$ the distance between the source and the detector~\cite{Akimov:2017ade}.
The $\nu_\mu$ component is produced from the stopped pion decays, $\pi^+\to \mu^++\nu_\mu$, with a monoenergetic flux at $(m^2_{\pi}-m_\mu^2)/(2 m_\pi)\simeq 30$~MeV. The $\bar{\nu}_\mu$ and $\nu_e$ components are produced from the subsequent muon decays, $\mu^+\to e^++\bar{\nu}_\mu+\nu_e$, with a kinematic upper bound at $m_\mu/2 \simeq 53$ MeV.
The presence of $\chi$-neutrino interaction will modify the COHERENT spectrum, which can be seen in Fig.~\ref{fig:spectrum}. We select two benchmark points to illustrate the effects of modified spectra:
\begin{itemize}
	\item Case A: $m_\chi=10$ MeV, $m_{Z^\prime}=100$ MeV, and $g_\chi g_q=5.0\times10^{-8}$\;,
	\item Case B: $m_\chi=100$ MeV, $m_{Z^\prime}=1000$ MeV, and $g_\chi g_q=5.0\times10^{-4}$\;.
\end{itemize}
Here we assume $g_{\chi L}=g_{\chi R}=g_{\chi}$. For these two sets of parameters, we fix $m_{Z'}=10 m_\chi$ for illustration and choose either $m_\chi<53$ MeV or $m_\chi>53$ MeV to exhibit the two regions where the loop contribution is significant or not.
In case A, the modification to the SM spectrum is dominated by the tree-level scattering process, $\nu N\to \chi N$, and the loop-level contribution can be neglected due to the small coupling constants. In case B, the tree-level process is kinematically forbidden for $m_\chi\gtrsim 53$ MeV, and the modification to the SM spectrum is only induced by the loop-level diagrams.

\section{Numerical results}
\label{sec:Res}

\subsection{Constraints}

The simplified neutrino model can be constrained by the invisible rare decays such as $K^+\to \pi^+ + {\rm invisible}$ via flavor changing neutral currents~\cite{Dolan:2014ska}. This rare decay process is recently measured by the NA62 experiment at CERN~\cite{CortinaGil:2020vlo}.
The actual calculation of the decay rate would suffer from a problem of UV divergence due to the fact that the simplified model here is not gauge-invariant~\cite{Dolan:2014ska,Arcadi:2017wqi}. The reliable estimate of the flavor observable relies on the UV completion realization.
As the $K\to \pi$ transitions are induced by flavor changing neutral currents, they set stringent constraints on $g_q$ coupling for up-type quarks~\cite{Hurtado:2020vlj,Li:2020pfy}.
We thus assume non-universal $g_q$ couplings for up-type and down-type quarks and neglect the up-type quark coupling in the following calculations. The UV problem does not affect our assumption of the couplings below and the corresponding conclusions.

Another constraint on the model is the effective number of  relativistic neutrino species $N_{eff}$ in the early Universe, where $N_{eff}$ is defined by
\begin{eqnarray}
N_{eff} ={8\over 7} \left({11\over 4}\right)^3 \left( {\rho_{rad} -\rho_\gamma \over \rho_\gamma }\right)
\end{eqnarray}
with $\rho_{rad}$ and $\rho_\gamma$ being the total radiation and photon energy densities, respectively. Considering the neutrino decoupling in the minimal SM, one has $N_{eff}^{\rm SM}=3.043\sim 3.045$~\cite{Bennett:2019ewm,EscuderoAbenza:2020qhd,Bennett:2020zkv,Akita:2020szl,Froustey:2020mcq}. Deviation from the SM prediction can be measured by CMB observations and the CMB Stage IV experiments are expected to reach a precision of $\Delta N =N_{eff}^{}-N_{eff}^{\rm SM}\sim 0.03$~\cite{CMB-S4:2016xur,Abazajian:2019eic} in the future. In our model, the thermal history of active neutrinos might be modified by the portal-like interaction in Eq.~(\ref{lagrang}), depending on parameter settings of the new coupling and masses of new species. For $m_{\chi}, ~m_{Z^\prime} \gg {\cal O}(1) ~{\rm MeV}$, new species already decay away at the time of neutrino decoupling, and any deviation from the $N_{eff}^{\rm SM}$ will be washed-out by weak interactions since neutrinos are still in thermal equilibrium with photon and electron at time of decays. Alternatively, if $m_{\chi},~m_{Z^\prime} \leq {\cal O}(1) ~{\rm MeV}$, $N_{eff}^{\rm SM}$ can be modified by the new interaction.
For $m_{Z^\prime}< m_{\chi}$, $Z^\prime$ can be taken as dark radiation if it is superlight, or it can directly decay into SM radiations, both of which contribute to the $N_{eff}^{}$. This effect has been widely studied by $X$ boson in explaining the Hubble tension problem~\cite{Dror:2019dib,Escudero:2019gzq,DiValentino:2021izs}.
For $m_\chi< m_{Z^\prime}$, $\chi$ may be produced in the early Universe either by neutrino oscillation via the Dodelson-Widrow mechanism~\cite{Dodelson:1993je} or by the annihilation of SM quarks. Then it decays into active neutrinos and photon, resulting in the deviation of $N_{eff}^{}$ from its SM value.
However, this parameter space is not favored by the CE$\nu$NS, as can be seen from the Fig.~\ref{fig:gchiL=R} in Sec.~\ref{sec:Res}, and $\chi$ is similar to a short-lived feebly interacting massive particle (FIMP)~\cite{Boyarsky:2021yoh} in this case.
One more interesting scenario is that $\chi$ is a long-lived particle and serves as a decaying dark matter. The Hubble tension problem can be partially solved~\cite{DiValentino:2021izs} in this case, but the contribution of new gauge interaction to CE$\nu$NS will be heavily suppressed because $g_\chi$ will be  negligibly small.
Since we mainly focus on CE$\nu$NS in this project, a systematic study of the impact of this model to $N_{eff}$ is beyond the reach of this paper and it will be presented in a future work.

\subsection{Results}

We evaluate the statistical significance of new physics beyond the SM by defining
\begin{align}
\chi^2 = \sum_{i=4}^{15} \left[\frac{N_\text{meas}^i-N_\text{th}^i(1+\alpha)-B_\text{on}(1+\beta)}{\sigma_\text{stat}^i}\right]^2+\left(\frac{\alpha}{\sigma_\alpha}\right)^2+\left(\frac{\beta}{\sigma_\beta}\right)^2\,,
\end{align}
where $N_\text{meas}^i$ and $N_\text{th}^i$ are the numbers of measured and predicted events per energy bin, respectively. Here $\alpha$ ($\beta$) is the nuisance parameter for the signal rate (the beam-on background) with an uncertainty of $\sigma_\alpha=0.28$ ($\sigma_\beta=0.25$)~\cite{Akimov:2017ade}.
The statistical uncertainty per energy bin is given by $\sigma_\text{stat}^i=\sqrt{ N_\text{meas}^i +2B_\text{SS}^i+B_\text{on}^i}$ with $B_\text{SS}^i$ being the steady-state background from the anti-coincident data, and $B_\text{on}^i$ the beam-on background mainly from prompt neutrons~\cite{COHERENT:2018imc}.

\begin{figure}[htb!]
\begin{center}
\includegraphics[scale=1,width=0.48\linewidth]{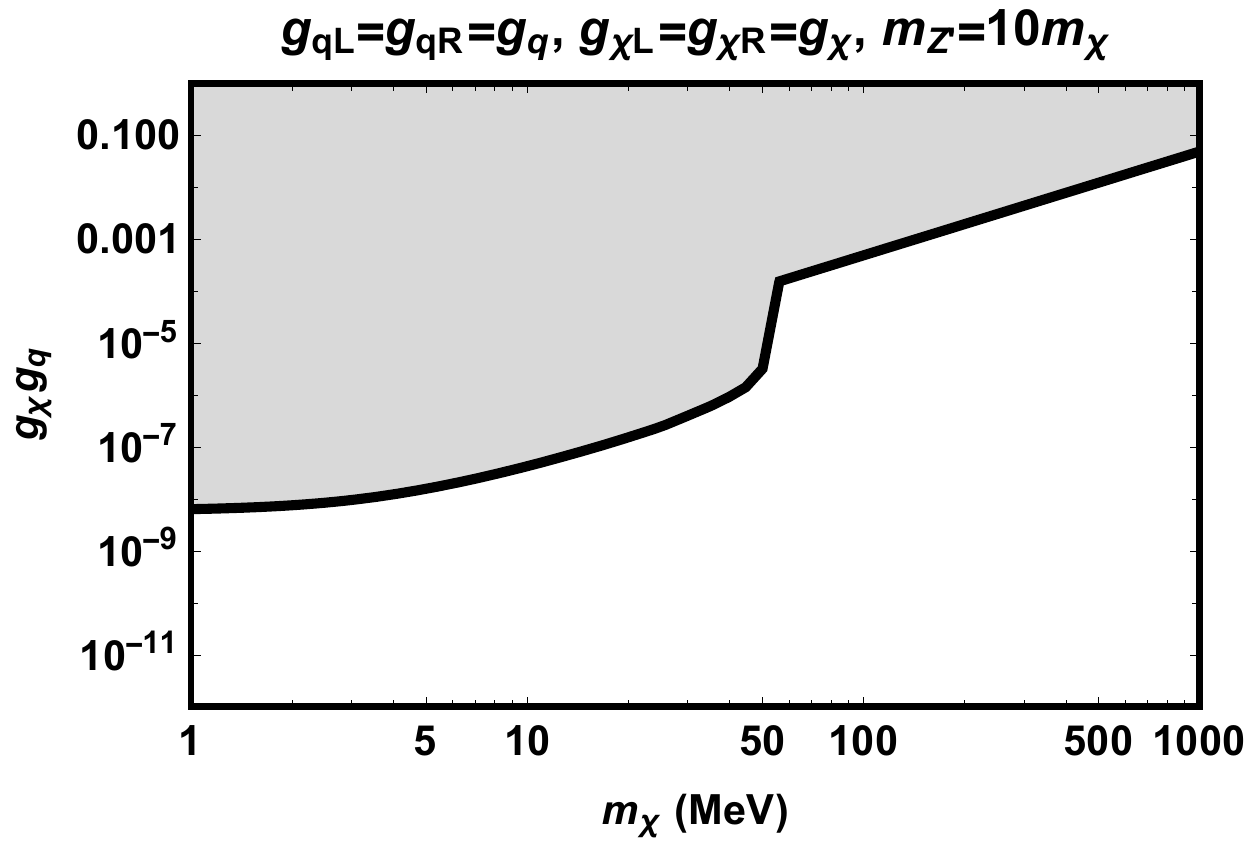}
\includegraphics[scale=1,width=0.48\linewidth]{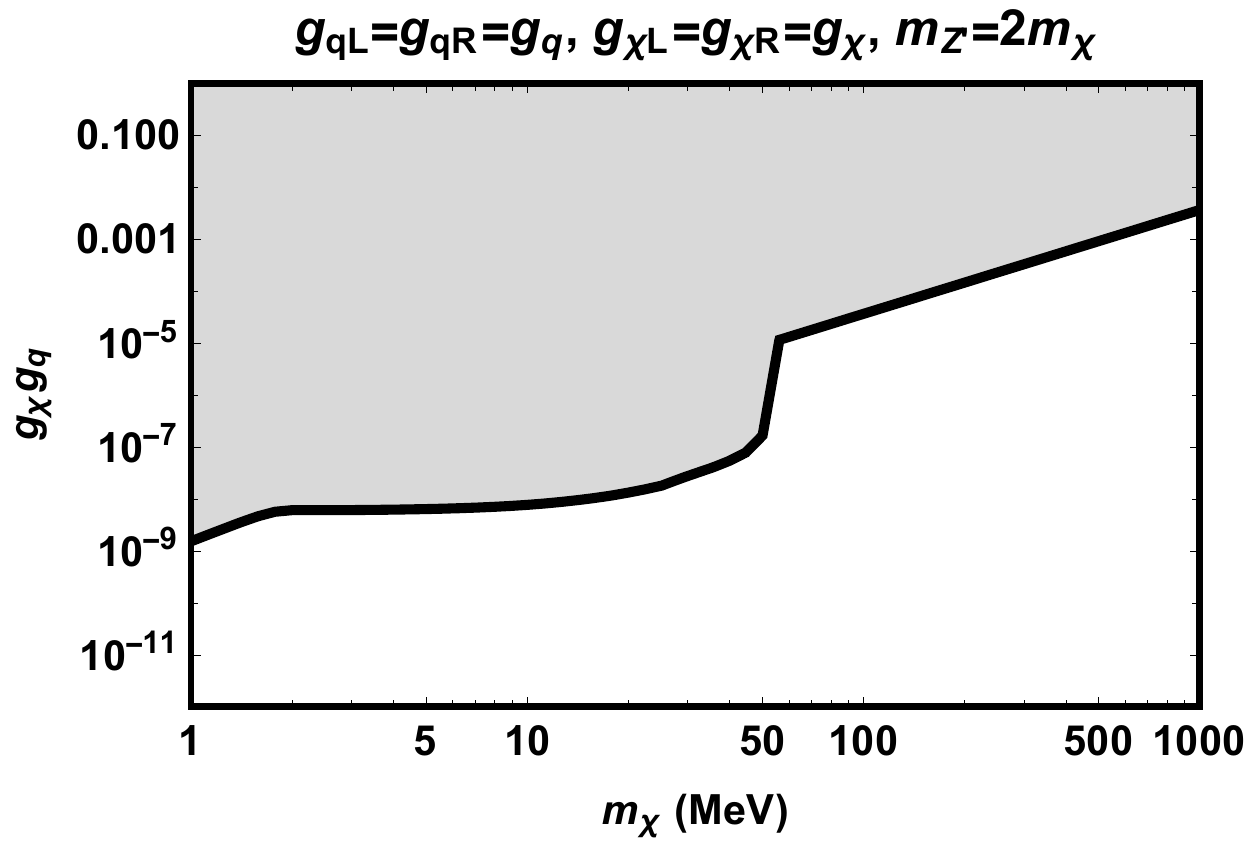}\\
\includegraphics[scale=1,width=0.48\linewidth]{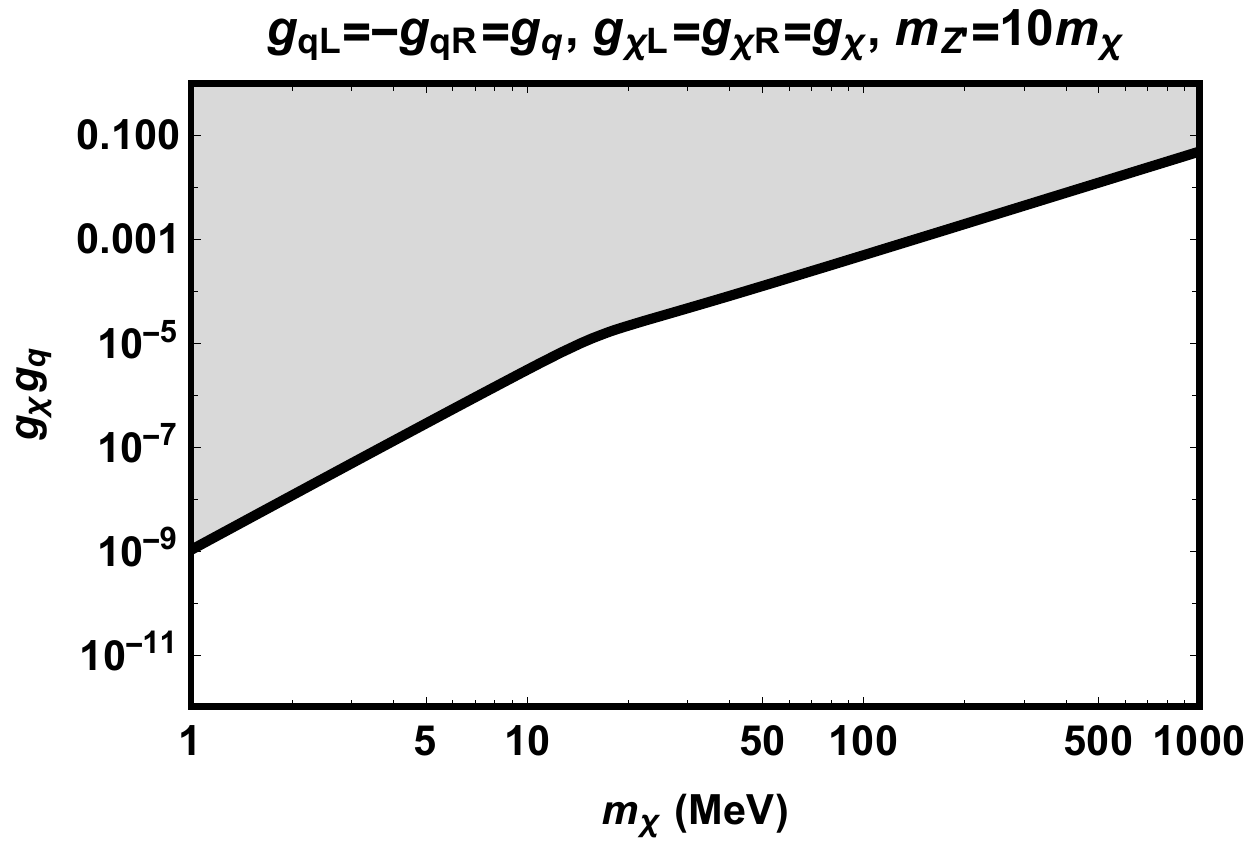}
\includegraphics[scale=1,width=0.48\linewidth]{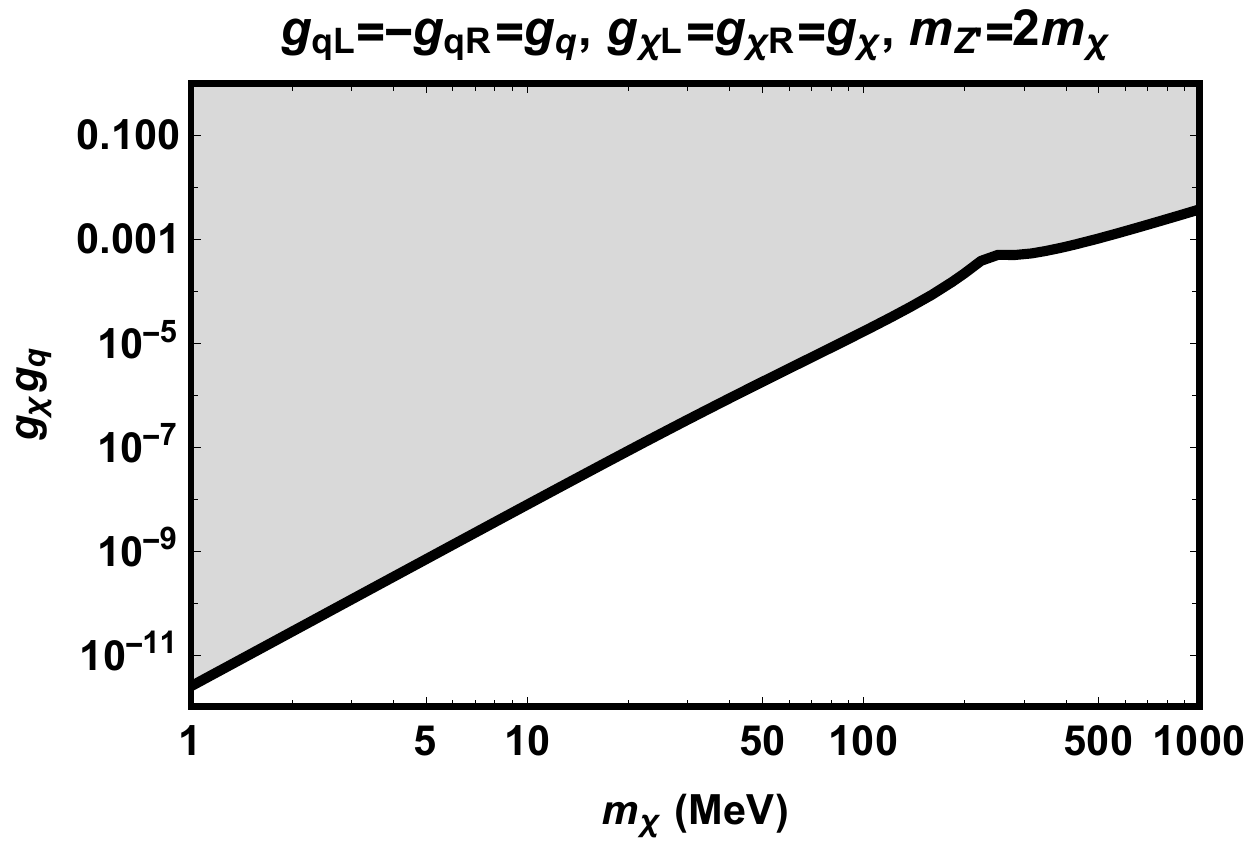}
\end{center}
\caption{The 90\% CL upper bounds on $g_\chi g_q$ as a function of the mass $m_\chi$ from COHERENT. We assume $g_{\chi L}=g_{\chi R}=g_\chi$, $g_{qL}=g_{qR}=g_q$ (top panels: case A) or $g_{qL}=-g_{qR}=g_q$ (bottom panels: case B) and $m_{Z'}=10 m_\chi$ (left), or $m_{Z'}=2 m_\chi$ (right).
}
\label{fig:gchiL=R}
\end{figure}

To obtain the bounds on the simplified neutrino model, we first set $g_{\chi L}=g_{\chi R}=g_\chi$ and $m_{Z'}=10m_\chi$ or $2m_\chi$ for both cases A and B, and scan over possible values of the product of the coefficients $g_\chi g_q$ for a given $m_\chi$. We choose one $Z'$ much heavier than $\chi$ and the other $Z'$ mass closer to $m_\chi$ for illustration. We expect the loop diagrams place more substantial contribution to the CE$\nu$NS process for the latter case. The 90\% CL upper bounds on $g_\chi g_q$ as a function of $m_\chi$ are shown in Fig.~\ref{fig:gchiL=R}. As we see from the top left panel in Fig.~\ref{fig:gchiL=R}, for $m_\chi \lesssim 53$ MeV in case A, the scattering is dominated by the tree-level process $\nu N\to \chi N$ and the upper bounds on $g_\chi g_q$ can reach as small as $6.7\times10^{-9}$ at $m_\chi= 1$ MeV for $m_{Z'}=10m_\chi$. The bounds become flat in small $m_\chi$ region, which can be understood from Eq.~(\ref{eq:treelevel}) since for small $m_\chi$ and $m_{Z^\prime}$ the tree-level process will be only sensitive to the coupling constants.
For $m_\chi\gtrsim 53$ MeV, however, the tree-level process $\nu N\to \chi N$ is kinematically forbidden and the relatively weaker bounds are entirely from the loop contributions. Thus, one can see a kink around $m_\chi\simeq 53$ MeV.
From the left panels of Fig.~\ref{fig:gchiL=R} to the right panels, in general the bounds become stronger as the mediator mass $m_{Z'}$ decreases. Also, from the top right panel in Fig.~\ref{fig:gchiL=R}, we see that for $m_{Z'}=2m_\chi$, the loop-level process becomes comparable to the tree-level process for small $m_\chi$, which gives a kink around $m_\chi\simeq 2$ MeV. The upper bounds on $g_\chi g_q$ can reach as small as $1.5\times10^{-9}$ at $m_\chi= 1$ MeV for $m_{Z'}=2m_\chi$.

We also show the 90\% CL upper bounds on $g_\chi g_q$ as a function of $m_\chi$ for case B in the bottom panels of Fig.~\ref{fig:gchiL=R}. The results are shown in the bottom left and right panel of Fig.~\ref{fig:gchiL=R} for $m_{Z'}=10m_\chi$ and $m_{Z'}=2m_\chi$, respectively. From Eq.~(\ref{eq:treeM}), one can see that the tree-level process in case B has no SI terms and the bounds for CE$\nu$NS are only determined by the loop-level contributions. Unlike case A with pure vector mediator $Z'$, the case B is induced by axial-vector current between $Z'$ and the SM quarks and has additional contributions from pNGB $\varphi$ in loop diagrams. The loop-level constraints on the couplings are thus stronger than those in case A.
For $m_\chi\simeq 53$ MeV in case B, the pure loop-level contribution constrains $g_\chi g_q$ at the level of $10^{-5}$ ($10^{-6}$) for $m_{Z'}=10m_\chi$ ($2m_\chi$). For small $m_\chi$ region, compared with the tree-level process in case A, the loop diagrams have stronger dependence on $m_\chi$ as expected.

\section{Conclusions}
\label{sec:Con}

We investigate the general neutrino interactions with an exotic fermion $\chi$ and a vector mediator in light of the coherent neutrino-nucleus scattering. We consider the framework of a simplified neutrino model in which a new Dirac fermion $\chi$ interacts with active neutrinos and a leptophobic vector mediator $Z'$. The chiral couplings between the mediator and the new fermion $\chi$ (the SM quarks) are parameterized by $g_{\chi L}$ and $g_{\chi R}$ ($g_{qL}$ and $g_{qR}$). At tree-level, the new fermion $\chi$ can be produced through the inelastic scattering process $\nu N\to \chi N$. We also include the loop-level contributions to the CE$\nu$NS process $\nu N\to \nu N$ with the new fermion $\chi$ running in the loop diagrams. For the choice of chiral couplings in the quark sector $g_{qL}=g_{qR}~(g_{qL}=-g_{qR})$, the CE$\nu$NS processes are dominated by the tree-level (loop-level) contribution. The COHERENT data are applied to place constraints on the couplings and the mass of fermion $\chi$.
We summarize our main conclusions in the following
\begin{itemize}
\item For the case of $g_{qL}=g_{qR}=g_q$ with vector current in quark sector, the scattering is mostly dominated by the tree-level process $\nu N\to \chi N$ for $m_\chi \lesssim 53$ MeV and the upper bounds on $g_\chi g_q$ can reach as small as $6.7\times10^{-9}$ ($1.5\times10^{-9}$ ) at $m_\chi= 1$ MeV for $m_{Z'}=10m_\chi$ ($2m_\chi$). For $m_\chi\gtrsim 53$ MeV, the bounds entirely come from the loop process $\nu N\to \nu N$ and are relatively weaker.
\item For the case of $g_{qL}=-g_{qR}$ with axial-vector current, the bounds for CE$\nu$NS are only induced by the loop-level contributions. For $m_\chi\simeq 53$ MeV in this case, the pure loop-level contribution constrains $g_\chi g_q$ at the level of $10^{-5}$ ($10^{-6}$) for $m_{Z'}=10m_\chi$ ($2m_\chi$).
\end{itemize}

\acknowledgments
TL is supported by the National Natural Science Foundation of China (Grant No. 11975129, 12035008) and ``the Fundamental Research Funds for the Central Universities'', Nankai University (Grant No. 63196013). JL is supported by the National Natural Science Foundation of China (Grant No. 11905299), Guangdong Basic and Applied Basic Research Foundation (Grant No. 2020A1515011479), the  Fundamental  Research  Funds  for  the  Central Universities, and the Sun Yat-Sen University Science Foundation. WC is supported by the National Natural Science Foundation of China under grant No. 11775025 and the Fundamental Research Funds for the Central Universities under grant No. 2017NT17.

\appendix

\section{Loop diagram calculation}

The Passarino-Veltman functions for the one-loop box diagrams are defined as
\begin{eqnarray}
D_0^b&\equiv& D_0[p_1^2,p_1^2,0,0,0,p_1^2;0,m_\chi^2,m_{Z'}^2,m_{Z'}^2]={-m_{Z'}^2+m_\chi^2+m_{Z'}^2{\rm ln}\Big({m_{Z'}^2\over m_\chi^2}\Big)\over m_{Z'}^2(m_{Z'}^2-m_{\chi}^2)^2}\;,\\
D_{00}^a-D_{00}^b&\equiv& D_{00}[p_1^2,p_1^2,0,0,0,p_1^2;0,m_\chi^2,0,m_{Z'}^2]-D_{00}[p_1^2,p_1^2,0,0,0,p_1^2;0,m_\chi^2,m_{Z'}^2,m_{Z'}^2]\nonumber \\
&=& {m_{Z'}^2-m_\chi^2-m_{Z'}^2{\rm ln}\Big({m_{Z'}^2\over m_\chi^2}\Big)\over 4(m_{Z'}^2-m_\chi^2)^2}\;,\\
D_{11}^a-D_{11}^b&\equiv& D_{11}[p_1^2,p_1^2,0,0,0,p_1^2;0,m_\chi^2,0,m_{Z'}^2]-D_{11}[p_1^2,p_1^2,0,0,0,p_1^2;0,m_\chi^2,m_{Z'}^2,m_{Z'}^2]\nonumber \\
&=& m_{Z'}^2{(m_{Z'}^2-m_\chi^2)(m_{Z'}^2+5m_\chi^2)-2m_\chi^2(2m_{Z'}^2+m_\chi^2){\rm ln}\Big({m_{Z'}^2\over m_\chi^2}\Big)\over 6m_\chi^2(m_{Z'}^2-m_\chi^2)^4}\;,\\
D_{001}^a-D_{001}^b&\equiv& D_{001}[p_1^2,p_1^2,0,0,0,p_1^2;0,m_\chi^2,0,m_{Z'}^2]-D_{001}[p_1^2,p_1^2,0,0,0,p_1^2;0,m_\chi^2,m_{Z'}^2,m_{Z'}^2]\nonumber \\
&=& m_{Z'}^2{-2m_{Z'}^2+2m_\chi^2+(m_{Z'}^2+m_\chi^2){\rm ln}\Big({m_{Z'}^2\over m_\chi^2}\Big)\over 12(m_{Z'}^2-m_\chi^2)^3}\;,\\
D_1^b&\equiv& D_1[p_1^2,p_1^2,0,0,0,p_1^2;0,m_\chi^2,m_{Z'}^2,m_{Z'}^2]={2(m_{Z'}^2-m_\chi^2)-(m_{Z'}^2+m_\chi^2){\rm ln}\Big({m_{Z'}^2\over m_\chi^2}\Big)\over 2(m_{Z'}^2-m_\chi^2)^3}\;,\\
D_{00}^b&\equiv& D_{00}[p_1^2,p_1^2,0,0,0,p_1^2;0,m_\chi^2,m_{Z'}^2,m_{Z'}^2]={-m_{Z'}^2+m_\chi^2+m_\chi^2{\rm ln}\Big({m_{Z'}^2\over m_\chi^2}\Big)\over 4(m_{Z'}^2-m_\chi^2)^2}\;,
\end{eqnarray}

The $F_G$ function for the two-loop diagrams with two pNGB $\varphi$ mediators in case B is
\begin{eqnarray}
F_{G}(p_1^2,m_\chi^2,m_{Z'}^2,m_q^2)&=& \int_0^1 dx\Big[-3{\partial\over \partial m_{Z'}^2}X_1\Big(p_1^2,m_\chi^2,m_{Z'}^2,{m_q^2\over x(1-x)}\Big)\nonumber \\
&+&{3m_q^2(1+x-x^2)\over x^2(1-x)^2}{\partial\over \partial m_{Z'}^2}X_2\Big(p_1^2,m_\chi^2,m_{Z'}^2,{m_q^2\over x(1-x)}\Big)\nonumber \\
&-&{4{m_q^4}(1-3x+3x^2)\over x^3(1-x)^3}{\partial\over \partial m_{Z'}^2}X_3\Big(p_1^2,m_\chi^2,m_{Z'}^2,{m_q^2\over x(1-x)}\Big) \Big] \;,
\end{eqnarray}
where
\begin{eqnarray}
X_1\Big(p_1^2,m_\chi^2,m_{Z'}^2,{m_q^2\over x(1-x)}\Big)&=& {1\over m_{Z'}^2-{m_q^2\over x(1-x)}}\Big[B_0(p_1^2,m_{Z'}^2,m_\chi^2)-B_0\Big(p_1^2,{m_q^2\over x(1-x)},m_\chi^2\Big)\Big]\;,\\
X_2\Big(p_1^2,m_\chi^2,m_{Z'}^2,{m_q^2\over x(1-x)}\Big)&=& {1\over m_{Z'}^2-{m_q^2\over x(1-x)}}\Big[X_1\Big(p_1^2,m_\chi^2,m_{Z'}^2,{m_q^2\over x(1-x)}\Big)-C_0\Big(p_1^2,{m_q^2\over x(1-x)},m_\chi^2\Big)\Big]\;,\nonumber \\ \\
X_3\Big(p_1^2,m_\chi^2,m_{Z'}^2,{m_q^2\over x(1-x)}\Big)&=& {1\over m_{Z'}^2-{m_q^2\over x(1-x)}}\Big[X_2\Big(p_1^2,m_\chi^2,m_{Z'}^2,{m_q^2\over x(1-x)}\Big)-D_0\Big(p_1^2,{m_q^2\over x(1-x)},m_\chi^2\Big)\Big]\;,\nonumber \\
\end{eqnarray}
and
\begin{eqnarray}
\int {d^4\ell\over (2\pi)^4}{1\over [(\ell+p)^2-M^2](\ell^2-m^2)}&=& {i\over (4\pi)^2}B_0(p^2,m^2,M^2)\;,\\
\int {d^4\ell\over (2\pi)^4}{1\over [(\ell+p)^2-M^2](\ell^2-m^2)^2}&=& {i\over (4\pi)^2}C_0(p^2,m^2,M^2)\;,\\
\int {d^4\ell\over (2\pi)^4}{1\over [(\ell+p)^2-M^2](\ell^2-m^2)^3}&=& {i\over (4\pi)^2}D_0(p^2,m^2,M^2)\;.
\end{eqnarray}

\bibliography{refs}

\end{document}